\documentclass[letterpaper]{llncs}

\usepackage{graphicx}
\usepackage{times}
\usepackage{listings}
\usepackage{hyperref}
\usepackage{color}
\usepackage{xspace}
\usepackage{subfigure}
\renewcommand{\subsubsection}{}
\renewcommand{\subparagraph}{}
\usepackage{titlesec}
\titlespacing{\section}{0pt}{2ex}{0.5ex}
\titlespacing{\subsection}{0pt}{1ex}{0.5ex}
\titlespacing{\subsubsection}{0pt}{0.5ex}{0ex}
\titlespacing{\paragraph}{0pt}{0.0ex}{0ex}

\newcommand{\runtime}{BDDT-SCC\xspace}

\begin{document}

\title{\runtime: A Task-parallel Runtime for Non Cache-Coherent Multicores}

\author{
  Alexandros Labrineas\inst{1}
  \and
  Polyvios Pratikakis\inst{1}
  \and
  Dimitrios S.~Nikolopoulos\inst{2}
  \and
  Angelos Bilas\inst{1}
}

\institute{
  Foundation for Research and Technology - Hellas
  \and
  Queens University of Belfast
}

\lstset{
  language=C,
  columns=flexible,
  numbers=left,
  showstringspaces=false,
  alsoletter={-},
  literate={-}{-}1,
  numbersep=1em,
  xleftmargin=2.5em,
  xrightmargin=1em,
  morecomment=[l][\bf\color{BrickRed}]{\#pragma\ scoop},
  commentstyle=\color{gray},
  keywordstyle=\bf\color{MidnightBlue},
  stringstyle=\color{OliveGreen},
}

\maketitle

\begin{abstract}

This paper presents \runtime, a task-parallel runtime system for non
cache-coherent multicore processors, implemented for the Intel
Single-Chip Cloud Computer. The \runtime runtime includes a dynamic
dependence analysis and automatic synchronization, and executes
OpenMP-Ss tasks on a non cache-coherent architecture. We design a
runtime that uses fast on-chip inter-core communication with small
messages. At the same time, we use non coherent shared memory to
avoid large core-to-core data transfers that would incur a high volume
of unnecessary copying. We evaluate \runtime on a set of
representative benchmarks, in terms of task granularity, locality, and
communication.  We find that memory locality and allocation plays a
very important role in performance, as the architecture of the SCC
memory controllers can create strong contention effects. We suggest
patterns that improve memory locality and thus the performance of
applications, and measure their impact.

\end{abstract}

\section{Introduction}

The rising core counts of modern processors trend towards hundreds of
cores in the near future.  However, the
performance of cache-coherent shared memory does not scale well with
the number of cores, leading to systems with high core counts that
have either expensive cache-coherent, non-uniform memory access
(cc-NUMA) or no cache-coherence at
all~\cite{scc,runnemede,lyberis:fccm12,lyberis:thesis}.
The Single-chip Cloud Computer (SCC) is a manycore processor that
represents this trend.  It consists of 48 cores, placed in a tile
formation with two cores per tile.  Tiles are connected by a
mesh, which also links with four memory controllers that address
the external system memory.  The memory address space can be either
private to each core or shared by all cores, although access to shared
memory is not cache coherent.  As there is no OS that
can currently use such a manycore processor, the SCC cores are
completely independent: each core runs an individual OS.

Programming such systems requires careful consideration of memory
allocation, layouts, locality and access patterns, as not all
memory accesses are equally expensive.  The common abstraction
of shared memory can greatly hurt performance and even break program
correctness (for non cache-coherent systems). More importantly, this
trend seems to continue strong in the future; recent work from Intel
predicts future manycores will not have fully coherent
caches~\cite{runnemede,last_millenium} and will require a change in
runtimes and operating system design.

Traditional threaded programming is not portable in future manycores.
Implicit communication between threads using shared memory does not
work through non-coherent memories and can hurt performance on cc-NUMA
memory. Moreover, clusters and systems like the SCC require explicit
communication among cores, which is complex for the programmer to
handle. For these reasons, the ``threads \& shared memory'' model is
not suitable for these systems. Conversely, task-parallel programming
models are better fit for such architectures because they lift the
effort required for explicit communication from the programmer to the
runtime system.

Task-based parallelism is expressed via annotations in the code that
identify certain procedure calls as concurrent tasks. This is a more
abstract way to express parallelism. The programmer describes all
parallelism without having to manually manage thread or process
communication and execution. The runtime extracts the best parallelism
automatically according to the system load and the available hardware
resources. 

Parallel programs require synchronization mechanisms to
produce correct executions. In early task parallel
systems~\cite{cilk,sequoia,openmp}, the programmer must use
such mechanisms to avoid conflicting memory accesses. Recent
task-parallel systems introduce implicit synchronization using
dependence analysis to order task execution and avoid
conflicts~\cite{tzenakis:appt13,pop:taco13,legion,pratikakis:mspc11,jenista:ppopp11}.  In contrast to
statically expressed parallelism, dynamic dependence analysis only
synchronizes tasks that actually have conflicting memory footprints
allowing the runtime to discover more parallelism.  However, existing
task-parallel runtimes target either shared-memory
multiprocessors~\cite{cilk,ompss,tzenakis:appt13} or clusters of
nodes that communicate over a network~\cite{sequoia,legion}, both very
different architectures to non-coherent manycores like the SCC.

Overall, this paper makes the following contributions:
\vspace{-0.5\baselineskip}
\begin{itemize}

  \item We design and implement \runtime, a task-parallel runtime
  system with implicit synchronization, for the SCC, a non-coherent
  manycore architecture.

  \item We evaluate \runtime using a set of representative benchmarks
  and find that memory contention and task granularity play a very
  important role in the scalability of the benchmarks.

\end{itemize}

\section{The SCC processor}

\begin{figure}[t]
\begin{center}
\includegraphics[width=0.9\linewidth]{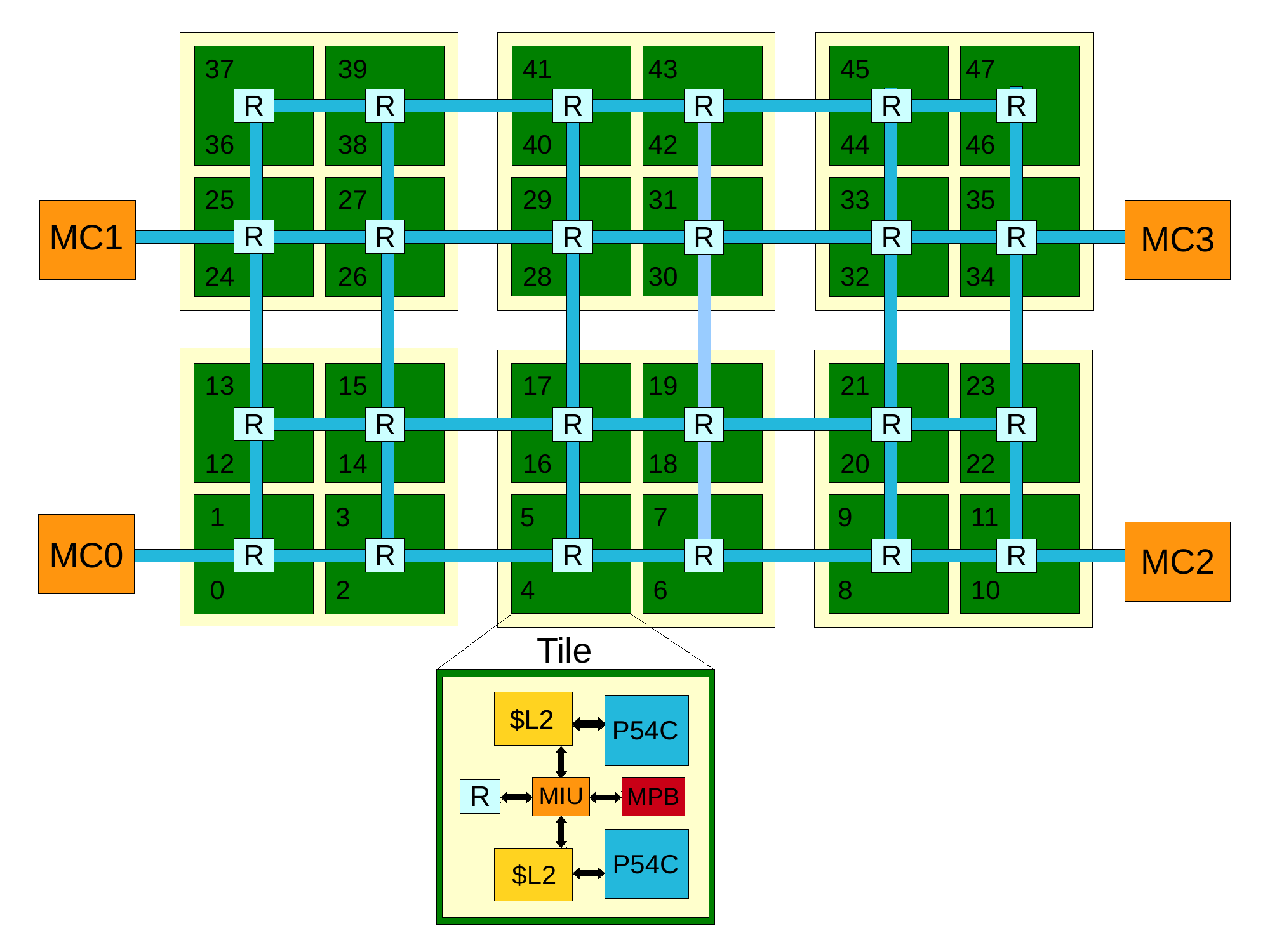}
\end{center}
\vspace{-5ex}
\caption{The SCC architecture}
\label{fig:arch}
\vspace{-3ex}
\end{figure}

Figure~\ref{fig:arch} shows the architecture of the SCC many-core
processor~\cite{scc-eas}. The SCC chip consists of 48 cores, placed in
a tile formation with two cores in each tile.  Each core has a unique
ID ranging from 0 to 47.  A 6$\times$4 mesh connects the tiles to each
other and to four memory controllers that address the external system
memory.  Each core has a private L1 instruction cache of 16KB, a
private L1 data cache of 16KB and a private unified L2 cache of 256KB.
Each dual-core tile has 16KB of SRAM dedicated to message passing.
This amounts to an on-chip \emph{message-passing buffer} (MPB) of 8KB
for each core.  The MPBs are memory-mapped and accessible from all
cores.  The chip features extensive frequency and voltage control on a
per tile and voltage island basis.  For all measurements in this paper
the cores are clocked at 533MHz, the mesh network at 800MHz and the
memory controllers at 800MHz.

The SCC processor uses four Memory Controllers (MCs) to address
off-chip memory.  The controllers addresses external DRAM using a
physical-to-physical translation through programmable LUTs.  By
default, each core gets a separate partition of the available DRAM and
runs a separate instance of the Linux kernel.  The on-chip LUTs can
also be programmed so that all cores in the SCC can physically address
a set of shared 16MB pages, split among the four memory
controllers~\cite{vanderWijngaart:osr11}.  These can be memory-mapped
by user-level processes running on separate cores so that they share
up to 512MB of DRAM.  However, the SCC does not implement cache
coherency, so it is the programmer's responsibility to flush the
write-combine buffers (equivalent to a write fence) and invalidate the
caches (equivalent to a read fence) of cores that access shared memory
so that written values become visible to readers
correctly.\footnote{Unfortunately, the 512MB shared-memory
configuration of the SCC overlaps some physical pages used by the
Linux kernel of four cores causing these kernels to panic.  We omit
the crashed Linux kernel cores from our benchmarks.}

\section{Design and Implementation}

%
\subsection{Programming model}

\runtime, like BDDT~\cite{tzenakis:appt13}, implements the OmpSs
programming model~\cite{ompss} for the SCC processor.  In OmpSs, the
programmer specifies function calls as tasks to be spawned using
compiler pragma directives.  A \emph{master} core executes the main
program and creates tasks to be executed in parallel by \emph{worker}
cores.  The programmer also specifies task footprints as memory
address ranges or multidimensional array tiles.  Every task argument
is described with a specific data access attribute, corresponding to
three access patterns: read (IN), write (OUT) and read/write (INOUT).
Dynamic analysis uses these attributes to discover task footprints
that overlap in memory and detects dependencies between tasks.

To detect dependencies, \runtime performs block-level dependence
analysis on the task arguments, similarly to BDDT.  The block-level
dependence analysis uses a custom allocator to split all allocated
memory into blocks and discovers task dependencies by detecting
whether any arguments of any two tasks contain the same block.

\begin{figure}[t]
\begin{center}
\includegraphics[width=0.95\textwidth]{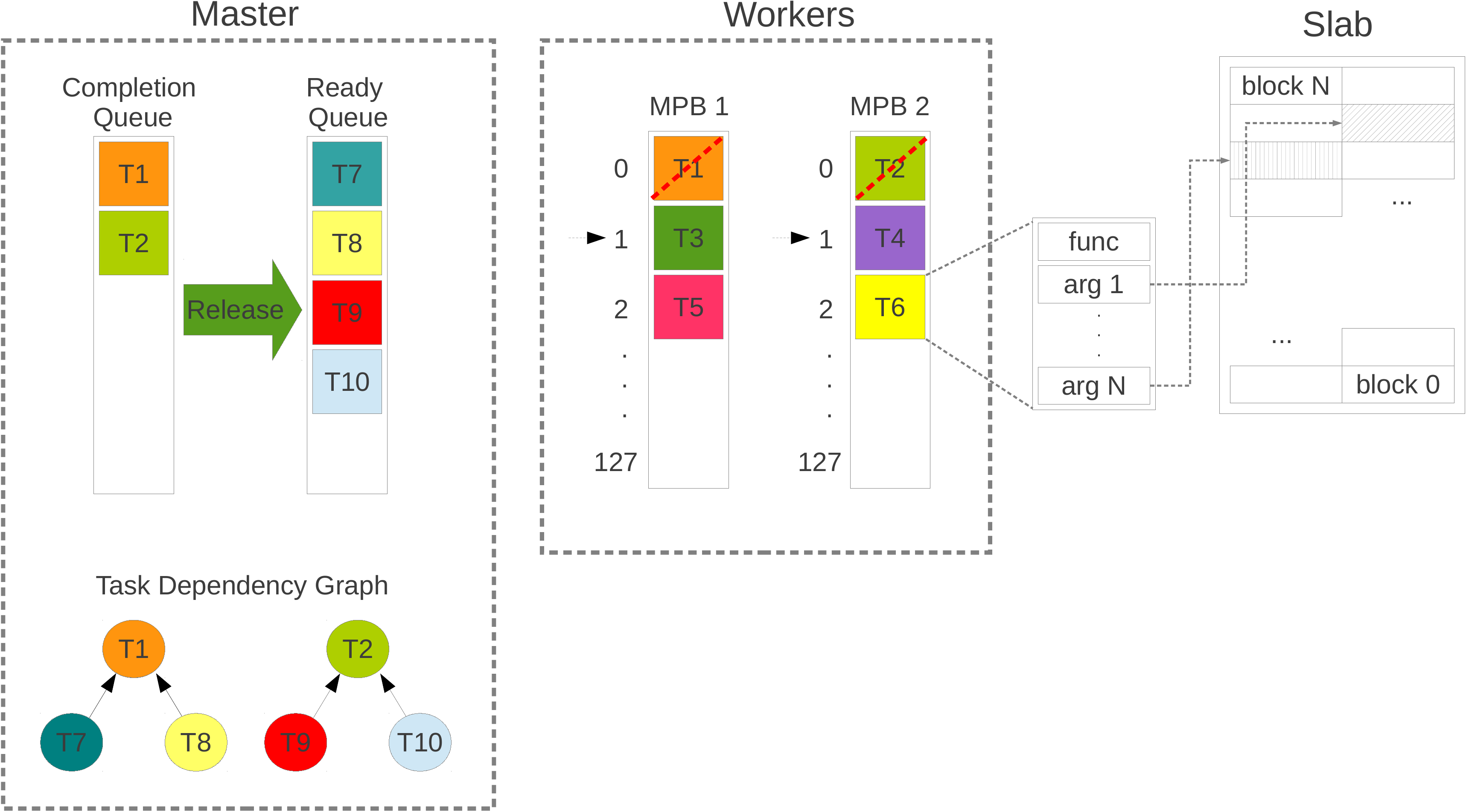}
\end{center}
\vspace{-5ex}
\caption{Example state during execution}
\label{fig:example}
\vspace{-3ex}
\end{figure}

Every task spawned by the application creates a new task instance
which goes through four stages in the runtime:
\vspace{-0.5\baselineskip}
\begin{itemize}

  \item Task initiation, in which the master creates a new task
  descriptor, detects its dependencies and either adds it to the task
  graph to wait for its arguments, or marks it as \emph{ready} to run
  if it has no dependencies.

  \item Task scheduling, in which the master assigns a ready task to
  an available worker.

  \item Task execution, in which the worker runs the task to
  completion on its specified arguments and marks it as complete.

  \item Task release, in which the runtime removes any dependencies on
  the completed task ---possibly creating new ready tasks--- and
  recycles its descriptor data.

\end{itemize}
\vspace{-0.5\baselineskip}
For example, the lower left part of Figure~\ref{fig:example} shows a
task dependency graph with two immediately ready tasks (T1 and T2) and
four dependent tasks (T7, T8, T9 and T10) that cannot run before T1
and T2 have completed.

\subsection{Memory Management}

Within \runtime, each instance of a spawned task corresponds to a
\emph{task descriptor}: a struct that includes a reference to the
spawned function, its arguments, and a representation of the task
footprint.  To keep track of tasks throughout their lifetime, \runtime
inserts each task descriptor into an appropriate data structure.  The
\emph{ready queue} of the master core contains descriptors of tasks
that are ready to run but have not been scheduled for execution to any
of the worker cores.  The \emph{completion queue} of the master core
contains task descriptors of executed tasks, whose dependencies are
not yet released.  The \emph{task graph} of the master core contains
descriptors of tasks with unresolved data dependencies, that cannot be
executed before these dependencies are released.  \runtime allocates
the ready queue, completion queue and task graph in the master core's
private memory.

In the example of Figure~\ref{fig:example}, tasks T1 and T2 are in the
completion queue of the master core after they are finished executing
on workers 1 and 2 and before their dependencies have been released.
When the dependencies of T1 are released, tasks T7 and T8 enter the
ready queue of the master.  When T2 is released, T9 and T10 enter the
ready queue of the master.

\runtime allocates a queue of task descriptors per worker core.  We
allocate a task queue as an array in each worker's Message-Passing
Buffer.  As all cores' MPBs are accessible from all other cores, the
master core writes directly to each worker's MPB to enqueue tasks to
that worker's queue or collect tasks that have finished executing.
Note that each MPB consists of 512 32-byte cache lines and writing a
single byte in an MPB will update all 32 bytes of that cache line.
So, we align task descriptors to MPB cache lines to avoid false
sharing among the master and worker cores.

For example, the middle part of Figure~\ref{fig:example} shows the
message-passing buffers of two worker cores.  In MPB of worker 1 the
master core has scheduled tasks T1, T3 and T5, whereas in MPB of
worker 2 the master core has scheduled T2, T4 and T6.  In the state
shown, when tasks T1 and T2 finish executing at workers 1 and 2, they
are marked as complete.  Then, the master core will collect T1 and T2
into its completion queue and reuse their position in MPBs 1 and 2.

Unlike task queues and runtime metadata, application data is often
much larger than the available on-chip memory.  This means that
core-to-core message passing of application data will result in
DRAM-to-core (cache misses), synchronous core-to-core communication,
and core-to-DRAM (cache evicts) communication\footnote{\ An early
version of the runtime used solely message passing; we found this
scenario caused unnecessary memory traffic, limiting performance.}
Instead, \runtime allocates all application data in SCC shared memory,
using a custom slab allocator.
For instance, the right part of Figure~\ref{fig:example} shows the
contents of a task descriptor.  Each argument of the task references
several blocks of data allocated in the shared memory.

\subsection{Task initiation}

To spawn a new task, \runtime allocates and initializes a
new task descriptor.  To avoid the overhead of allocating and
deallocating task descriptors and also improve the locality and cache
performance of the runtime system, \runtime uses a pre-allocated
memory pool of task descriptors and recycles deallocated tasks.  If
there are no free task descriptors, the master core blocks until a
task is complete.
After creating a new task descriptor, the BDDT dependence analysis
detects any data dependencies between the new task and previous
tasks~\cite{tzenakis:appt13}.  If the new task depends on existing
tasks that have not completed yet, its task descriptor is added to the
dependence graph to wait until all the dependencies are resolved.  If
there are no dependencies, then the task is \emph{ready} to run.

To detect dependencies, \runtime uses the BDDT block-based dependence
analysis.  In short, \runtime uses a custom allocator to split the
application memory into memory blocks and keeps metadata for each
block.  The runtime creates block metadata for each task that operates
on any given block and uses the metadata to order tasks that use the
same data.  When a task is first in this ordering for all the blocks
of its arguments, it has no dependencies and it is ready to run.  For
details on the BDDT block dependence analysis, we refer the reader to
the corresponding technical report~\cite{tzenakis:tr12}.

\subsection{Task scheduling}

The master core can be in one of two modes: (i) \emph{running}, or
(ii) \emph{polling}.  Initially in running mode, the master core
starts executing the main program and schedules spawned tasks that are
\emph{immediately ready} to worker cores.  To schedule a task to a
worker core in this mode, the master core tries to append the task to
the task queue in the worker's MPB\footnote{\ This communication is
asynchronous: the master core writes directly to the remote MPB
without blocking or interrupting the worker}.  To do that, the master
keeps a local index of the next available entry in the MPB queue for
each worker and checks the state of this entry.  If the entry is
empty, the master writes the task descriptor in that entry.  If the
entry holds a completed task, the master enqueues the completed task
in the completion queue and replaces it with the ready task.  If the
next available entry is full, the master adds the task to a local
queue of ready tasks and continues with main program execution.  This
way, the master never blocks at a spawn and will resume the
application execution until either all tasks are spawned and it
reaches a synchronization point, or it runs out of task descriptors.

At all points where the execution of the main program blocks, the
master enters the polling mode.  This can happen at synchronization
points, which include explicit barriers and the end of the main
program, or during task creation, if there are no available task
descriptors.  During its polling mode, the master performs three
functions: (i) It removes ready tasks from the ready queue, as long as
it is not empty, and schedules them; (ii) it polls the queue entries
of each worker to discover task descriptors marked as completed; and
(iii) it removes completed tasks from the completion queue and
releases their dependencies.

When on polling mode, scheduling is similar to that on running state.
The master appends the descriptor to the next available entry.
However, if this entry is full, the master does not return the task
back in the ready queue as during the running mode, but continues with
the next worker. If all worker queues are full the master dequeues a
completed task from its completion queue, \emph{releases} its
dependencies and then retries scheduling of the first task.

\subsection{Task execution}

To execute a scheduled task, the worker core reads the next ready task
descriptor from its local MPB and executes the task.  Task execution
is simply a call of the task function on the task arguments.  The task
arguments are allocated in the external shared memory and are thus
accessible by all cores.  Note that the shared memory is cacheable, but the
SCC caches are not coherent.  Thus, \runtime requires every worker
core to invalidate its L2 cache before task execution and flush it
after it, to make the task output visible to all subsequent
tasks running on other cores and maintain program correctness.

After the worker executes the task function, it marks the task
descriptor as completed in the worker's MPB buffer task queue and
continues with the next ready task in the queue.  To avoid a race
between the master and the worker on the MPB task queue of the worker,
we use L1 invalidation as a read barrier and flushing the
write-combine buffer as a write barrier.  Specifically, the worker
invalidates its L1 cache before polling each task entry in the queue,
and flushes its write-combine buffer after changing a task descriptor
from \emph{ready} to \emph{completed}.  Conversely, the master
invalidates its L1 cache before reading a worker's queue.  As an
optimization, the master does not flush its write-combine buffer after
putting a ready task in a worker's queue.  This may mean that the
worker will not observe the transition from \emph{completed} to
\emph{ready} or from \emph{empty} to \emph{ready} for that entry
immediately.  That is not an issue, however, since it can only cause
the worker to poll its queue again.

\subsection{Task release}

The master core locates completed tasks in workers' task queues during
scheduling of newer ready tasks, or during polling its mode.  To avoid
extending the critical path, the master core does not process the
completed tasks immediately, but collects them in its completed queue,
recycling that space in the workers' queues into new task descriptors.
When the master core idles because all worker queues are full, it
has reached a barrier, or it needs to recycle the task resources, it
iterates the completed queue and lazily releases the completed tasks'
dependencies.  Releasing a completed task decrements a dependency
counter for each of its dependent tasks.  If a counter reaches zero,
the master removes the newly ready task from the dependency graph and
marks it as ready to run.

\section{Evaluation}

\subsection{Core Placement}

\begin{figure}[t]
\begin{minipage}{0.48\textwidth}
  \includegraphics[width=\textwidth]{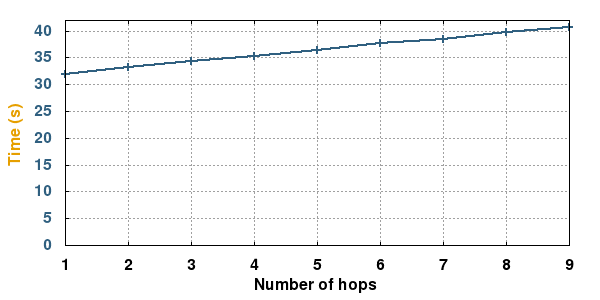}
  \vspace{-5ex}
  \caption{Memory access latency}
  \label{fig:latencies}
\end{minipage}
\begin{minipage}{0.48\textwidth}
  \includegraphics[width=\textwidth]{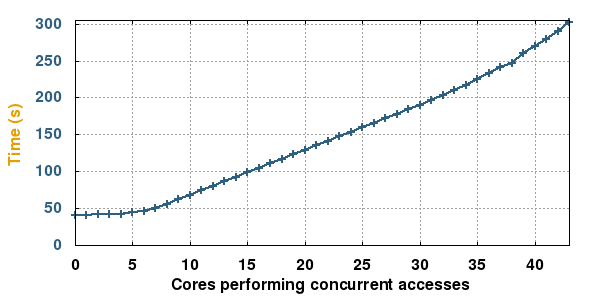}
  \vspace{-5ex}
  \caption{Memory contention effects}
  \label{fig:contention}
\end{minipage}
\vspace{-3ex}
\end{figure}

The SCC architecture results in a different latency for accessing DRAM
depending on a core's distance from the respective memory
controller~\cite{scc-guide,mattson:sc10}.  We measured the impact of
the latency difference using a microbenchmark that repeatedly accesses
a 16MB array allocated to take exactly one shared memory page managed
by controller 0.  Figure~\ref{fig:latencies} shows the total execution
time depending on how many hops away the core running the
microbenchmark was from the controller. Similarly, the MPB access
latency varies depending on the core's distance from the respective
MPB.

We took the variable latency into account when placing the cores in
\runtime, so that (i) the master core is one of the middle cores,
having almost uniform distance from all memory controllers and worker
cores, and (ii) each worker core is placed as close as possible to the
master.  Therefore, every additional worker has higher communication
cost with the master and its distance from the memory controllers
deviates from uniform.  For instance, a configuration with 31 workers
uses all the cores of one with 30 workers, plus an
additional core that is as close to the master as possible.

We placed the master at core 16, one of the middle cores on the SCC
(cores 16, 17, 18 and 19 in Figure~\ref{fig:arch}).  This position
minimizes the maximum distance to worker cores to 5 hops and the sum
of hops from the master to all remote MPBs at full chip utilization to
120 hops.  Similarly, the closest memory controller is 4 hops away
from the master and the furthest is 5 hops away.  The total distance
from the master core to all memory controllers is 18 hops. Placing the
master at any other position results into a higher number of total hops and
increases the runtime's communication overhead.


We found that the latency for accessing DRAM increases with the number
of cores performing concurrent accesses.  We use the same
microbenchmark to measure the impact of concurrent memory
accesses through the same memory controller.
Figure~\ref{fig:contention} shows the total execution time (y-axis) of
the microbenchmark run on a reference core, while the same
microbenchmark is running on various other cores (x-axis).  We select
the reference core to be the most distant (9 hops) from the memory
controller 0 as a worst-case scenario.  The total execution time on
the reference core increases with the number of accessing cores, due
to contention effects at the memory controllers.  This effect does not
occur at the default configuration of the SCC, where each core has
access to a disjoint part of the physical memory, accessible only via
the nearest memory controller.  With shared-memory, however,
contention effects become more pronounced, as all cores access all
memory controllers.

\subsection{Benchmarks}

\begin{figure}[!t]
  \subfigure[Black-Scholes]{
    \includegraphics[width=0.48\textwidth]{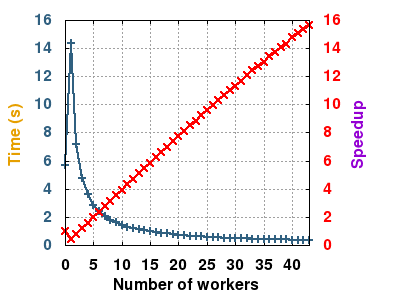}
    \label{fig:blsch}
  }
  \subfigure[Matrix Multiply]{
    \includegraphics[width=0.48\textwidth]{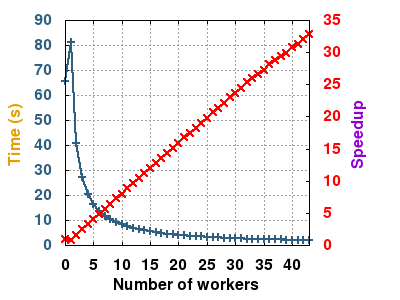}
    \label{fig:matmul}
  }

  \subfigure[Fast Fourier Transform]{
    \includegraphics[width=0.48\textwidth]{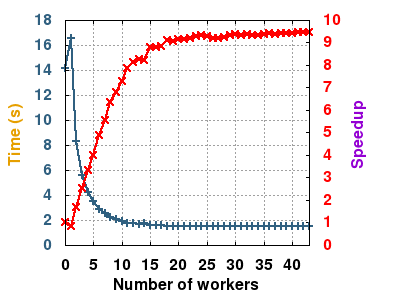}
    \label{fig:fft}
  }
  \subfigure[Jacobi]{
    \includegraphics[width=0.48\textwidth]{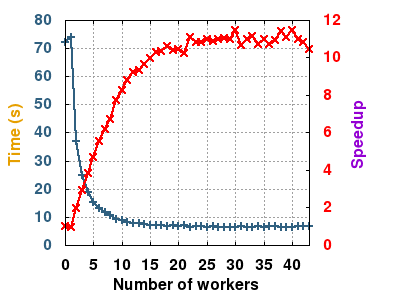}
    \label{fig:jacobi}
  }

  \subfigure[Cholesky]{
    \includegraphics[width=0.48\textwidth]{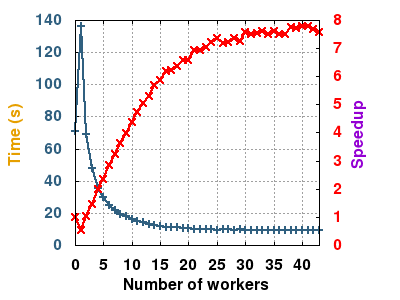}
    \label{fig:cholesky}
  }
  \hfil
  \subfigure{
    \fbox{\includegraphics[width=0.2\textwidth]{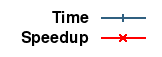}}
  }
  \hfill

\caption{Benchmark execution time and speedup}
\label{fig:results}
\vspace{-4ex}
\end{figure}

We use 5 well-known applications to evaluate the runtime.
\emph{Black-Scholes} is a financial application; we use a data set of
2M options, split into tasks of 512 options.
\emph{Matrix Multiply} is a tiled parallel implementation of matrix
multiplication; we used 1K$\times$1K floats, split into 64$\times$64
tiles.
\emph{Fast Fourier Transform} computes the FFT of a 2D matrix; we used
1M complex doubles, split into blocks of 32 rows at the
transformation phase and 32$\times$32 tiles at the transposition phase.
\emph{Jacobi Method} computes a Jacobian determinant; we used
4K$\times$4K floats, split into 512$\times$512 tiles, for 16
iterations.
\emph{Cholesky Decomposition} computes a matrix factorization; we used
2K$\times$2K doubles, split into 128$\times$128 tiles.
All applications except for \emph{Black-Scholes} have task dependencies.
Some benchmarks have small, concentrated datasets that fit within the
shared-memory segment of a single memory controller.  This creates
strong contention effects when all cores access memory through the
same memory controller.  In these cases, we use padding and non-unit
strides during allocation, to distribute application data across all
memory controllers as uniformly as possible.
%

\subsection{Results}

Figure~\ref{fig:results} shows the execution time (left y-axis) and
scalability (right y-axis) for each benchmark.  The x-axis shows the
number of worker cores used (\emph{i.e.}, we do not count the master
core).  We show the performance of the original, sequential program at
point 0.  The sequential program runs at the master core and allocates
all its memory at the nearest memory controller.  We exclude
initialization time from all measurements and report total time of
parallel execution.  Black-Scholes and Matrix Multiply scale to
16$\times$ and 33$\times$ speedups, respectively, compared to the
sequential execution.

\begin{figure}[!t]
  \subfigure[Black-Scholes]{
    \includegraphics[width=0.48\textwidth]{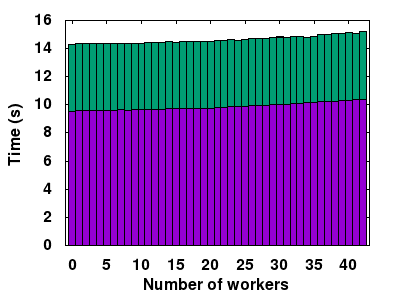}
    \label{fig:bd:blsch}
  }
  \subfigure[Matrix Multiply]{
    \includegraphics[width=0.48\textwidth]{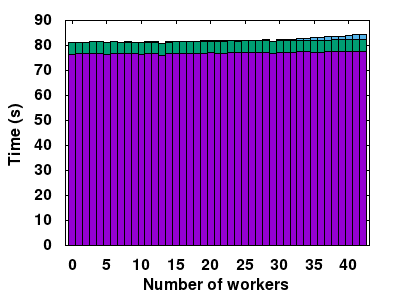}
    \label{fig:bd:matmul}
  }

  \subfigure[Fast Fourier Transform]{
    \includegraphics[width=0.48\textwidth]{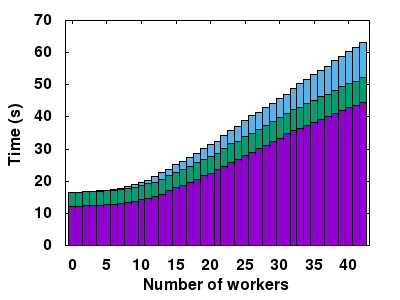}
    \label{fig:bd:fft}
  }
  \subfigure[Jacobi]{
    \includegraphics[width=0.48\textwidth]{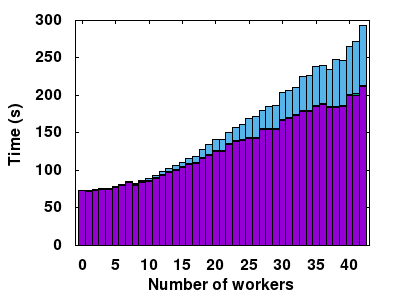}
    \label{fig:bd:jacobi}
  }

  \subfigure[Cholesky]{
    \includegraphics[width=0.48\textwidth]{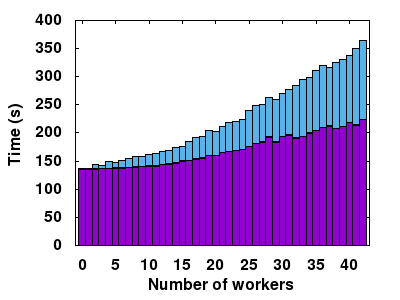}
    \label{fig:bd:cholesky}
  }
  \hfil
  \subfigure{
    \fbox{\includegraphics[width=0.2\textwidth]{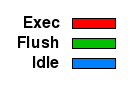}}
  }

\caption{Breakdown of total processor time per benchmark}
\label{fig:breakdown}
\vspace{-3ex}
\end{figure}

For each application, we present execution time breakdowns for
the worker cores. We break down the execution of each worker in
three parts: (i) time waiting the master (idle), (ii) time spent in
application code, and (iii) time spent for L2 cache flush and
invalidation. Figure~\ref{fig:breakdown} shows the cumulative
breakdowns for all participating cores. FFT, Jacobi and Cholesky
feature strong memory contention effects. The cumulative time spent in
application code grows as core count increases, since each individual
memory access or cache miss costs more.

\begin{figure}[!t]
  \subfigure[Black-Scholes]{
    \includegraphics[width=0.48\textwidth]{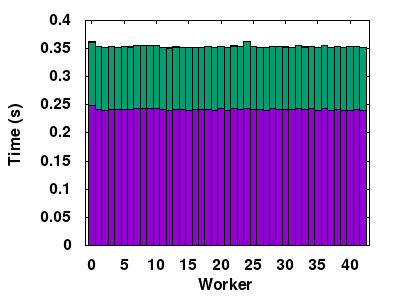}
    \label{fig:bal:blsch}
  }
  \subfigure[Matrix Multiply]{
    \includegraphics[width=0.48\textwidth]{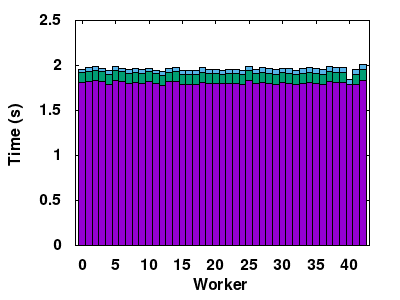}
    \label{fig:bal:matmul}
  }

  \subfigure[Fast Fourier Transform]{
    \includegraphics[width=0.48\textwidth]{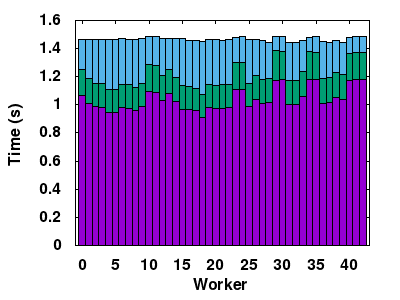}
    \label{fig:bal:fft}
  }
  \subfigure[Jacobi]{
    \includegraphics[width=0.48\textwidth]{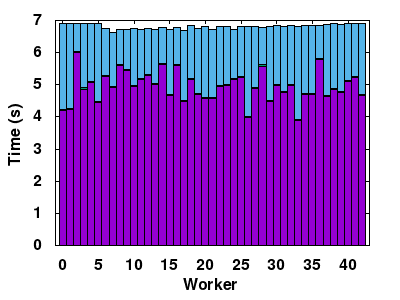}
    \label{fig:bal:jacobi}
  }
  
  \subfigure[Cholesky]{
    \includegraphics[width=0.48\textwidth]{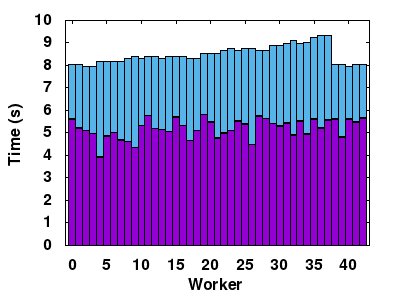}
    \label{fig:bal:cholesky}
  }
  \hfil
  \subfigure{
    \fbox{\includegraphics[width=0.2\textwidth]{breakdown-legend}}
  }

\caption{Load Balance for 43 workers}
\label{fig:balance}
\vspace{-3ex}
\end{figure}

Figure~\ref{fig:balance} shows the load balance per worker for the
configuration with 43 workers for each benchmark.  Again, we show the
breakdown of the total time spent by each worker, into: (i) time
waiting the master (idle), (ii) time spent in application code, and
(iii) time spent for L2 cache flush and invalidation.  Note that idle
time in the workers is always caused by too fine a task granularity,
where the master cannot spawn and schedule tasks fast enough to keep
all workers busy.

Black-Scholes scales linearly to all the available worker cores
(Figure~\ref{fig:blsch}).  However, its speedup is not equal to the
number of cores, due to the high flush time to execution time ratio
(Figure~\ref{fig:bd:blsch}).  In this case the master core is idle
most of its time, waiting for the workers to finish.  Black-Scholes
also produces a very balanced schedule, where all workers perform an
almost equal amount of work (Figure~\ref{fig:bal:blsch}.

Similarly, Matrix Multiply scales proportionally to the number of
workers (Figure~\ref{fig:matmul}), achieving better speedup than
Black-Scholes as the constant overhead of cache flushing is minimal
compared to execution time (Figure~\ref{fig:bd:matmul}).  Matrix
Multiply is also very well balanced, with all workers performing an
almost equal amount of work.
On the other hand, the rest of the applications do not exhibit similar
scalability.  In all cases, their scalability is limited by the memory
contention effects shown above.

FFT scales to 16 worker cores (Figure~\ref{fig:fft}), with
performance being almost unaffected by a larger number of workers.
Figure~\ref{fig:bd:fft} demonstrates the effect of contention: the
total execution time increases with the number of workers because
memory accesses become more expensive, although the total actual work
remains the same.  The flush time is also slightly affected by the
same contention effect, although the number of flushes is constant,
equal to the total number of tasks.  Also, the total idle time
starts increasing when the number of workers reaches 10, indicating
that the master core is not fast enough to serve all workers beyond
that point.  This also affects load balancing
(Figure~\ref{fig:bal:fft}), as the master core cannot keep up with
workers that execute faster tasks.

Similarly, Jacobi and Cholesky reach a maximum speedup at 22 worker
cores (Figures~\ref{fig:jacobi} and~\ref{fig:cholesky}).  Again, this
is a combination of memory contention that increases the total task
execution time with the number of workers, although the total work
remains the same (Figures~\ref{fig:bd:jacobi}
and~\ref{fig:bd:cholesky}, and also of the master becoming a
bottleneck at 13 and 3 worker cores, respectively, increasing the
worker's idle time.  In turn, this reduce the load balancing of the
problem for high core counts (Figures~\ref{fig:bal:jacobi}
and~\ref{fig:bal:cholesky}, as the master cannot schedule more tasks
fast to workers that finish early.

\section{Related Work}

Task-parallel programming models have risen as a high-level
alternative to thread programming.  Task-parallelism allows the
programmer to specify scoped regions of atomic code without
specifying synchronization or communication during a task.  Early
task-parallel programming models do not perform dependence analysis or
implicit synchronization~\cite{openmp,cilk,tbb,sequoia}.  Recent
task-parallel systems add either static~\cite{best:pldi11,jenista:ppopp11} or dynamic dependence
analysis~\cite{perez:ics10,smpss,legion}.  Most
systems target shared memory architectures, where
cache-coherency automates most of the communication, or clusters,
where everything is communicated through message-passing.  In
comparison, \runtime targets the SCC, a manycore processor without
cache coherency or asynchronous message-passing
communication, although all cores can share physical memory.

The SCC processor relaxes hardware cache-coherence to improve
scalability and energy consumption~\cite{scc,mpronline-scc,scc-guide,scc-eas}.  Early runtimes treat the SCC as a
message-passing system~\cite{vanderWijngaart:osr11,mpi-scc}, use
distributed and cluster languages to program it~\cite{x10-scc}, or
implement software cache-coherence~\cite{dsm-scc}.  However, these
approaches fail to take advantage of the non-coherent shared memory of
the SCC and also the granularity of tasks that does not require
coherence traffic for individual loads and stores.

\section{Conclusions}


We present \runtime, a runtime system for executing
task-parallel programs written in the OmpSs programming model on the
SCC.  We demonstrate that \runtime scales up to a factor of 33$\times$
in applications where the memory traffic is balanced over the four
memory controllers of the chip and the tasks' footprint features good
cache locality (\emph{i.e.,} Matrix Multiply).  We found that
applications with dense, stencil computations reach a scalability
limit due to strong memory contention effects before taking full
advantage of the chip.
We conclude that task-parallel programs can take advantage of non
cache coherent architectures such as the SCC manycore processor
through careful consideration of locality and data placement, and load
balancing of data across memory controllers.
We project that scalability could be greatly improved by
(i) a mechanism for asynchronous bulk communication between
processors, lifting the limit of 8KB through-MPB messages;
(ii) hardware support for fine-grained management of the cache,
reducing the amount of cache misses and consequently contention
effects: the SCC uses an older P54C core that does not support L2
partial flushing or separate invalidation.
Overall, the SCC performs better on data-parallel applications
and coarse-grained parallel programs.  Although fine-grained
parallelism has greater potential for speed-up, in our current design,
a too-fine granularity could make scheduling tasks the bottleneck,
limiting scalability.


\bibliographystyle{splncs}
\bibliography{polyvios}

\begin{thebibliography}{10}

\bibitem{scc}
Howard, J., et~al.:
\newblock A 48-core ia-32 message-passing processor with {DVFS} in 45nm cmos.
\newblock In: Solid-State Circuits Conference Digest of Technical Papers
  (ISSCC). (2010)  108--109

\bibitem{runnemede}
Carter, N.P., Agrawal, A., Borkar, S., Cledat, R., David, H., Dunning, D.,
  Fryman, J., Ganev, I., Golliver, R.A., Knauerhase, R., Lethin, R., Meister,
  B., Mishra, A.K., Pinfold, W.R., Teller, J., Torrellas, J., Vasilache, N.,
  Venkatesh, G., Xu, J.:
\newblock Runnemede: An architecture for ubiquitous high-performance computing.
\newblock In: HPCA. (2013)

\bibitem{lyberis:fccm12}
Lyberis, S., Kalokairinos, G., Lygerakis, M., Papaefstathiou, V., Tsaliagkos,
  D., Katevenis, M., Pnevmatikatos, D., Nikolopoulos, D.:
\newblock Formic: Cost-efficient and scalable prototyping of manycore
  architectures.
\newblock In: FCCM. (2012)

\bibitem{lyberis:thesis}
Lyberis, S.:
\newblock Myrmics: A Scalable Runtime System for Global Address Spaces.
\newblock PhD thesis, University of Crete (August 2013)

\bibitem{last_millenium}
Knauerhase, R., Cledat, R., Teller, J.:
\newblock For extreme parallelism, your os is sooooo last-millennium.
\newblock In: HotPar. (2012)

\bibitem{cilk}
Blumofe, R.D., Joerg, C.F., Kuszmaul, B.C., Leiserson, C.E., Randall, K.H.,
  Zhou, Y.:
\newblock Cilk: an efficient multithreaded runtime system.
\newblock In: PPoPP. (1995)

\bibitem{sequoia}
:
\newblock The sequoia programming language.
\newblock \url{http://http://sequoia.stanford.edu}

\bibitem{openmp}
Dagum, L., Menon, R.:
\newblock {OpenMP}: An industry-standard {API} for shared-memory programming.
\newblock IEEE Comput. Sci. Eng. \textbf{5} (January 1998)

\bibitem{tzenakis:appt13}
Tzenakis, G., Papatriantafyllou, A., Vandierendonck, H., Pratikakis, P.,
  Nikolopoulos, D.:
\newblock {BDDT}: Block-level dynamic dependence analysis for task-based
  parallelism.
\newblock In: APPT. Lecture Notes in Computer Science (2013)

\bibitem{pop:taco13}
Pop, A., Cohen, A.:
\newblock {OpenStream}: Expressiveness and data-flow compilation of {OpenMP}
  streaming programs.
\newblock TACO \textbf{9}(4) (January 2013)  53:1--53:25

\bibitem{legion}
Bauer, M., Treichler, S., Slaughter, E., Aiken, A.:
\newblock Legion: expressing locality and independence with logical regions.
\newblock In: SC. (2012)

\bibitem{pratikakis:mspc11}
Pratikakis, P., Vandierendonck, H., Lyberis, S., Nikolopoulos, D.S.:
\newblock A programming model for deterministic task parallelism.
\newblock In: MSPC. (2011)

\bibitem{jenista:ppopp11}
Jenista, J.C., Eom, Y.H., Demsky, B.:
\newblock {OoOJava}: Software out-of-order execution.
\newblock In: PPoPP. (2011)

\bibitem{ompss}
Duran, A., Ayguade, E., Badia, R.M., Labarta, J., Martinell, L., Martorell, X.,
  Planas, J.:
\newblock {OmpSs}: a proposal for programming heterogeneous multi-core
  architectures.
\newblock Parallel Processing Letters \textbf{21}(02) (2011)  173--193

\bibitem{scc-eas}
{Intel Labs}:
\newblock {SCC} external architecture specification (2010)

\bibitem{vanderWijngaart:osr11}
van~der Wijngaart, R.F., Mattson, T.G., Haas, W.:
\newblock Light-weight communications on intel's single-chip cloud computer
  processor.
\newblock SIGOPS Oper. Syst. Rev. \textbf{45}(1) (February 2011)  73--83

\bibitem{tzenakis:tr12}
Tzenakis, G., Papatriantafyllou, A., Zakkak, F., Vandierendonck, H.,
  Pratikakis, P., Nikolopoulos, D.S.:
\newblock {BDDT}: Block-level dynamic dependence analysis for deterministic
  task-based parallelism.
\newblock Tech Report 426, {FORTH} (February 2012)

\bibitem{scc-guide}
{Intel Labs}:
\newblock The {SCC} programmer's guide (2012)

\bibitem{mattson:sc10}
Mattson, T.G., Riepen, M., Lehnig, T., Brett, P., Haas, W., Kennedy, P.,
  Howard, J., Vangal, S., Borkar, N., Ruhl, G., Dighe, S.:
\newblock The 48-core scc processor: the programmer's view.
\newblock In: SC. (2010)

\bibitem{tbb}
Reinders, J.:
\newblock Intel threading building blocks. First edn.
\newblock O'Reilly \& Associates, Inc., Sebastopol, CA, USA (2007)

\bibitem{best:pldi11}
Best, M.J., Mottishaw, S., Mustard, C., Roth, M., Fedorova, A., Brownsword, A.:
\newblock Synchronization via scheduling: Techniques for efficiently managing
  shared state.
\newblock In: PLDI. (2011)

\bibitem{perez:ics10}
P{\'e}rez, J.M., Badia, R.M., Labarta, J.:
\newblock Handling task dependencies under strided and aliased references.
\newblock In: International Conference on Supercomputing. (2010)

\bibitem{smpss}
:
\newblock SMP Superscalar ({SMPSs}) v2.3 User's Manual. (2010)

\bibitem{mpronline-scc}
Baron, M.:
\newblock The single-chip cloud computer

\bibitem{mpi-scc}
Ure\~{n}a, I.A.C., Riepen, M., Konow, M., Gerndt, M.:
\newblock Invasive mpi on intel's single-chip cloud computer.
\newblock In: International Conference on Architecture of Computing Systems.
  ARCS'12 (2012)

\bibitem{x10-scc}
Chapman, K., Hussein, A., Hosking, A.L.:
\newblock X10 on the single-chip cloud computer: porting and preliminary
  performance.
\newblock In: ACM SIGPLAN X10 Workshop. X10'11 (2011)

\bibitem{dsm-scc}
Kim, J., Seo, S., Lee, J.:
\newblock An efficient software shared virtual memory for the single-chip cloud
  computer.
\newblock In: Asia-Pacific Workshop on Systems. APSys'11 (2011)

\end{thebibliography}

\end{document}